\documentclass[journal]{IEEEtran}
\usepackage{cite}
\usepackage{amsmath,amssymb,amsfonts}
\usepackage{algorithmic}
\usepackage{graphicx}
\usepackage{textcomp}
\usepackage{multirow}
\usepackage{url}
\usepackage{hyperref}
\usepackage{tabularx}
\usepackage{multirow}
\usepackage{array}
\usepackage{booktabs}
\usepackage{hhline}

\begin{document}
\title{Comparative Analysis of ImageNet Pre-Trained Deep Learning Models and DINOv2 in Medical Imaging Classification}
\author{Yuning Huang, Jingchen Zou, Lanxi Meng, Xin Yue, Qing Zhao, Jianqiang Li, Changwei Song, Gabriel Jimenez, Shaowu Li, Guanghui Fu\textsuperscript{*}
\thanks{Jingchen Zou, Yuning Huang, Xin Yue, Qing Zhao, Jianqiang Li, Changwei Song are with School of Software Engineering, Beijing University of Technology, Beijing, China.}
\thanks{Lanxi Meng, Shaowu Li are with Department of Neuroimaging, Beijing Neurosurgical Institute, Beijing Tiantan Hospital, Capital Medical University, Beijing, China.}
\thanks{Guanghui Fu and Gabriel Jimenez are with Sorbonne Universit\'{e}, Institut du Cerveau – Paris Brain Institute - ICM, CNRS, Inria, Inserm, AP-HP, H\^{o}pital de la Piti\'{e}-Salp\^{e}tri\`{e}re, Paris, France (e-mail: guanghui.fu@inria.fr).}
\thanks{Corresponding author: Guanghui Fu (\url{guanghui.fu@inria.fr})}
\thanks{This work has been submitted to the IEEE for possible publication. Copyright may be transferred without notice, after which this version may no longer be accessible.}
}
\maketitle
\begin{abstract}
Medical image analysis frequently encounters data scarcity challenges. Transfer learning has been effective in addressing this issue while conserving computational resources. The recent advent of foundational models like the DINOv2, which uses the vision transformer architecture, has opened new opportunities in the field and gathered significant interest. However, DINOv2's performance on clinical data still needs to be verified. In this paper, we performed a glioma grading task using three clinical modalities of brain MRI data. We compared the performance of various pre-trained deep learning models, including those based on ImageNet and DINOv2, in a transfer learning context. Our focus was on understanding the impact of the freezing mechanism on performance. We also validated our findings on three other types of public datasets: chest radiography, fundus radiography, and dermoscopy. Our findings indicate that in our clinical dataset, DINOv2's performance was not as strong as ImageNet-based pre-trained models, whereas in public datasets, DINOv2 generally outperformed other models, especially when using the frozen mechanism. Similar performance was observed with various sizes of DINOv2 models across different tasks. In summary, DINOv2 is viable for medical image classification tasks, particularly with data resembling natural images. However, its effectiveness may vary with data that significantly differs from natural images such as MRI. In addition, employing smaller versions of the model can be adequate for medical task, offering resource-saving benefits. 
Our codes are available at \url{https://github.com/GuanghuiFU/medical_dino_eval}. 

\end{abstract}
\IEEEpeerreviewmaketitle

\section{Introduction}

\IEEEPARstart{D}{eep} learning has made significant strides, attracting the attention of researchers and finding applications across a broad range of fields~\cite{lecun2015deep}. The emergence of large-scale datasets, such as ImageNet~\cite{deng2009imagenet}, has been a significant driver of this growth, fueling the development of more sophisticated models~\cite{huh2016makes}. However, in medical image analysis, the lack of large datasets poses a significant challenge~\cite{tajbakhsh2020embracing}. This shortage is primarily due to the high costs of data annotation and strict patient privacy laws. In response, researchers are investigating pre-feature learning methods for medical image analysis tasks. Among these methods, transfer learning is particularly prominent as it utilizes extensive datasets to initially train models, which are subsequently fine-tuned for specific medical imaging tasks~\cite{zhuang2020comprehensive, morid2021scoping}. Simultaneously, self-supervised learning, which derives meaningful features from unlabeled data, is emerging as a valuable approach in medical image analysis~\cite{krishnan2022self}. Though transfer and self-supervised learning both focus on feature learning, they apply different strategies in data use and learning techniques, underscoring the adaptability of deep learning in overcoming the distinct challenges of medical image analysis.

Recently, foundation models have had widespread attention. Models like OpenAI's ChatGPT~\cite{brown2020language} have excelled in natural language processing tasks~\cite{thirunavukarasu2023large, he2023towards}, highlighting the potential for general artificial intelligence~\cite{moor2023foundation, rishi2021opportunities}. In image analysis, the rising prominence of foundation models in computer vision has introduced novel approaches to the field~\cite{huang2023visual, mazurowski2023segment}. Notable examples include SAM~\cite{sam_kirillov2023segment}, focused on image segmentation, and DINOv2~\cite{oquab2023DINOv2}, a versatile backbone model trained on extensive datasets for a range of applications, such as image classification, retrieval, depth estimation, and semantic segmentation.
The advancement of foundational models in computer vision has led medical researchers to consider their application in medical image analysis. For instance, Yue et al.~\cite{yue2023morphology} used SAM and heat maps generated by a classification model to segment breast cancer images. Huix et al.~\cite{huix2024natural} explored the use of DINOv2 in medical image classification tasks, finding that it outperforms ImageNet pre-trained models in effectiveness. However, the need for more clinical data evaluation points to further investigation into these models' internal mechanisms.
Additionally, the nature and effectiveness of transfer learning and foundation models raise questions about whether foundation models consistently surpass pre-trained deep learning models. It necessitates extensive research across various tasks to ascertain their comparative advantages.

Glioma, the most common primary intracranial tumor, originates from neuroglial stem or progenitor cells~\cite{weller2015glioma}. The World Health Organization's 2021 classification of central nervous system tumors categorizes brain gliomas into grades 1 to 4~\cite{komori2022grading}. In the United States, the annual incidence rate of brain glioma is 6 per 100,000 people, with a 5-year mortality rate of approximately 5\%~\cite{ostrom2018adult}. Developing a computer model capable of swiftly classifying brain MRIs could significantly aid physicians in quickly assessing the condition and planning further treatment~\cite{kv2019glioma}. Consequently, developing relevant deep learning models is vital~\cite{buchlak2021machine}. More studies applying these models to clinical data are necessary to determine their suitability for clinical use~\cite{jin2024evaluating}.

In this study, we utilized clinical data to compare the performance of foundation models and ImageNet pre-trained deep learning models in three-modality brain MRI for glioma grading. We specifically focused on the impact of the freezing mechanism in training, where specific layers of the model are kept fixed while others are fine-tuned. Similar experiments were conducted on three public datasets, involving tasks like chest X-ray classification\cite{chestxray_kermany2018identifying}, glaucoma classification on eye fundus\cite{eye_orlando2020refuge}, and pigmented skin lesion detection on dermatoscopic images\cite{skin_tschandl2018ham10000}. We observed that the freezing approach with DINOv2 consistently yielded superior results compared to training without freezing. However, in our clinical trials, DINOv2 did not surpass the pre-trained deep learning models in performance, although it excelled in public datasets. This study also extends to clinical applications, exploring MRI's potential in tumor grading.

\section{Related work} \label{sec:related} 
\subsection{Brain image classification and transfer learning}
Deep learning techniques have been widely used in medical image classification tasks~\cite{litjens2017survey}.
Fu et al.~\cite{fu2021attention} developed an interpretable model that utilizes an attention mechanism specifically designed to identify cerebral hemorrhage in brain CT scans. This model highlights suspicious areas using an attention heatmap, providing a basis for its predictions and assisting medical professionals. 
Following this, Wang et al.~\cite{wang2022diagnosis} expanded on this approach by focusing the attention mechanism more intensely on image methods to enhance feature learning. This refined approach, which involves the fusion of these methods, resulted in improved performance. In human-computer interaction experiments involving doctors diagnosing the same images, it was observed that each set of images could be processed 10 seconds faster, coupled with a significant increase of 16 points in the F1 score. These results underscore the efficacy of artificial intelligence algorithms in medical diagnostics.
Soleymanifard et al.~\cite{soleymanifard2022multi} proposed a multi-stage tumor classification method for a complete tumor classification process. This study utilized a 12-scale multiscale fuzzy C-mean (MsFCM) detection of enhanced tumors. After segmenting the tumors and normal tissues, the input tumors were classified into low-grade gliomas (LGG) and high-grade gliomas (HGG) using a fully connected neural network. However, this study was not tested on clinical data and only validated the method on public data sets. 
Kaur et al.,~\cite{kaur2020deep} investigated applying several pre-trained deep learning models to classify pathological brain images. This research primarily concentrated on assessing the performance of diverse Deep Convolutional Neural Network (DCNN) architectures, including AlexNet, ResNet, GoogLeNet, and VGG, using datasets from the Harvard Repository, a clinical dataset, and Figshare's multi-class tumor dataset. Moreover, the research enhanced classification efficiency by optimizing the final three layers of these models.

These studies highlight the potential and applicability of deep learning technology in brain image analysis research. However, it is important to note that most of these studies rely on public datasets. In contrast, clinical data often presents more challenges due to its complexity and scarcity. Transfer learning emerges as a valuable solution to these issues. It effectively addresses the challenges of data scarcity and also conserves time and computational resources, offering the potential for enhanced performance in these contexts~\cite{kim2022transfer, yu2022transfer}.
Deepak et al.,~\cite{deepak2019brain} extracted features of brain tumor images using GoogLeNet and then performed five-fold cross-validation experiments on images using SVM as a classifier on the Figshare dataset to classify brain tumor types. The performance was evaluated using the area under the curve (AUC), precision, recall, F-score, and specificity. The results demonstrate that transfer learning is a valuable technique in scenarios with limited availability of medical images.
Chelghoum et al.~\cite{chelghoum2020transfer} employed deep transfer learning to classify three types of brain tumors using a CE-MRI dataset. They utilized nine deep CNN architectures as feature extractors, modifying the network's last three layers for fine-tuning. Additionally, they investigated the impact of epoch variations on classification performance.
With its training on large-scale datasets, the foundation model offers flexible adaptability for various downstream tasks, presenting new opportunities in the context of transfer learning. The specific performance and principles of the two are still worth exploring.

\subsection{Foundation model for image classification}
The deep learning field is increasingly applying foundation models that can be adapted for diverse tasks. A notable example in computer vision is the Segment Anything model (SAM)~\cite{sam_kirillov2023segment}, a foundation model that has gathered significant interest, particularly in medical image segmentation tasks~\cite{sam_med_review_huang2023segment, sam_med_review_mazurowski2023segment}. Another representative model is DINOv2~\cite{caron2021emerging}, a label-free, self-distillation approach leveraging the Vision Transformer architecture, notable for outperforming other self-supervised and some supervised learning methods. Building on DINOv2's success, Oquab et al.~\cite{oquab2023DINOv2} developed DINOv2, an enhanced version incorporating elements from iBOT~\cite{zhou2021ibot}, and pre-trained on the LVD-142 M dataset with 142 million natural images. DINOv2, maintaining DINOv2's strengths and adding further improvements, is adaptable for tasks like classification, segmentation, depth estimation, and image retrieval in both image and video formats.
Based on these benefits from the foundation model, researchers are working on applying DINOv2 to medical image analysis. Baharoon et al.~\cite{baharoon2023towards} used the DINOv2 model for radiological image analysis and conducted over 100 experiments using DINOv2 on different modalities (X-ray, CT, and MRI) with tasks including disease classification and organ segmentation. The results of this study show that DINOv2 has good cross-task generalization capabilities relative to supervised, self-supervised, and weakly supervised models. However, the researchers maintained a frozen backbone throughout their experiment, a strategy of considerable importance in transfer learning. The underlying mechanism of this approach warrants further investigation due to its significance. 
Prokop et al.~\cite{prokop2023deep} extensively evaluated models such as DINOv1 and DINOv2, among others, using various pre-trained CNN and vision transformers (ViTs) as feature extractors and combinations of multiple classifiers in a small-sample X-ray image classification scenario. The research tasks include the COVID-19 recognition task and the tuberculosis recognition task. In both tasks, ViTs as feature extractors outperform CNN-based models in almost all scenarios. This suggests that DINOv2 and MAE visual transformers may be a good choice as feature extractors in metric learning models. However, the mean AUROC of DINOv2-ViT-B/14 lagged behind that of DINOv2-ViT-B/8 in the COVID-19 recognition task. 
Huix et al.~\cite{huix2024natural} conducted a validation of five foundation models across four medical image classification datasets. Their research revealed that using DINOv2 as a backbone for transfer learning exhibits strong transferability, potentially supplanting the role of ImageNet pre-training in medical classification tasks. Additionally, their study delves into the freezing mechanism, observing that freezing the foundation model frequently resulted in a decline in performance in several datasets.

These studies show the potential and effectiveness of the DINOv2 model, highlighting its significant promise for applications in medical image analysis. However, it is important to note that most of these studies rely on experiments using public data and lack validation with clinical data, which is crucial for comprehensive assessment and practical applicability in medical settings.

\section{Methods}
In this research, we conducted a comparative analysis of the ImageNet pre-trained deep learning and foundation models using a private clinical dataset and three additional public datasets. The study emphasized examining the influence of the model's freezing mechanism on performance outcomes. We carried out six targeted experiments on two distinct datasets. Moreover, the study specifically explored the models' ability to grade gliomas within clinical datasets, with comprehensive results detailed in the subsequent results section. 
\subsection{Transfer learning on ImageNet pre-trained models}
We selected several representative deep-learning models pre-trained on ImageNet for comparison, including VGG16, ResNet50, and DenseNet121. These models were selected for their excellent performance in large-scale image recognition tasks.

\subsubsection{VGG16} 
The VGG network~\cite{simonyan2014very} marked a significant advancement in deep learning for image classification. VGG16, in particular, represented a major step forward in computational vision with its 16-layer architecture, consisting of 13 convolutional and 3 fully connected layers. This depth, combined with small (3x3) convolution filters, significantly enhanced the model's capability for intricate feature extraction. The design of VGG16 was instrumental in enabling the identification of complex and abstract visual patterns, setting a new standard for deep network architectures in image processing. The application of VGG16 in medical imaging, mainly through transfer learning, has shown significant efficacy~\cite{hossain2023breast, gayathri2023exploring, veni2023vgg, islam2023fine}. 

\subsubsection{ResNet50} 
ResNet~\cite{HeZRS15} has significantly advanced the field of image classification. Its key innovation lies in the approach to Residual Learning, which effectively tackles the vanishing gradient problem common in deep neural networks. ResNet achieves this through the integration of 'skip connections,' which allow a portion of the input to bypass one or more layers and then merge with the output of subsequent layers. This architectural feature ensures the network can learn identity functions where necessary, maintaining training stability and improving learning efficiency. Importantly, this design enables the successful training of much deeper networks, leading to marked improvements in performance on complex image classification tasks. In the field of medical imaging, ResNet50 has been successfully applied to disease classification through transfer learning~\cite{sharma2023brain, athisayamani2023feature, gouda2020skin, mehnatkesh2023intelligent}.

\subsubsection{DenseNet121} 
DenseNet~\cite{HuangLW16a} represents a breakthrough in the architecture of convolutional networks, particularly with its densely connected design. Unlike traditional architectures, each layer in DenseNet receives inputs from all preceding layers, ensuring comprehensive feature integration. This structure significantly enhances the flow of information and gradients throughout the network, leading to more effective training and reduced information loss. A standout aspect of DenseNet is its exceptional efficiency in feature reuse, which allows for a reduced number of parameters without sacrificing performance. Among its variants, DenseNet121 is particularly notable for its robust performance in various image classification tasks, especially in processing images with complex textures and patterns. Its application in medical image analysis, particularly in disease diagnosis and tumor detection, has been extensively recognized and adopted~\cite{girdhar2023densenet, lanjewar2023lung, prakash2023densenet, sonawane2023skin}.


\subsection{DINOv2: vision transformer-based computer vision foundation model}
In the realm of image data processing, DINOv2~\cite{oquab2023DINOv2} outperforms traditional CNNs in handling large-scale, complex datasets. Distinguishing itself from other Transformer-based models, DINOv2's unique strength lies in its capacity for self-supervised learning, enabling it to effectively train on unlabeled datasets. This model demonstrates proficiency in a variety of image processing applications, such as image classification, object detection, and segmentation, attributed to its robust feature extraction capabilities. Researchers are increasingly investigating the use of DINOv2 for medical image analysis tasks~\cite{huix2024natural, prokop2023deep, baharoon2023towards}.



\section{Datasets}

\subsection{Clinical datasets: glioma grading in brain MRI}
In this study, we studied glioma images from 101 patients. 
The study was approved by the Ethics Committee of Beijing Tiantan Hospital, affiliated with Capital Medical University (Project No. KY 2020-128-01). 
Clinical data of patients with histopathologic diagnosis of glioma were collected from Picture Archiving and Communication Systems (PACS), and 101 patients were finally enrolled after screening by inclusion and exclusion process. The inclusion criteria included: 1) patients with biopsy or surgical resection followed by pathological tissue confirmation of glioma, with the tumor located on the cerebellar vermis; 2) patients with no history of therapeutic surgeries such as hormone, radiotherapy, or puncture before MRI scanning; 3) MRIs were acquired preoperatively; and 4) the presence of a clear pathologic diagnosis. On the other hand, the exclusion criteria included 1) incomplete preoperative images or unclear pathological organization and 2) poor quality MRIs (motion artifacts, metal artifacts, others). 
According to the 2016 revised World Health Organization (WHO) classification of central nervous system (CNS) tumors\cite{komori20172016}: diffuse gliomas include WHO grade II and III astrocytomas, grade II and III oligodendrocytes Cytomas, grade IV glioblastoma and childhood-related diffuse gliomas. This study divided clinical MRI glioma data into 3 categories based on histological and pathological diagnosis: grade II, grade III and grade IV.

The acquisition equipment and scanning protocols are described as follows: Imaging was conducted using a Siemens 3T Prisma MRI, complemented by a 20-channel head and neck coil. Initial localization was achieved via a preliminary imaging scan. After axial level determination on the sagittal plane for the trans anterior union, a series of structural image scans were executed. These included:

\begin{itemize}
\item T1-Weighted 3D Structural Imaging: Parameters were set at TR=2300ms, TE=2.32ms, flip angle = 8°, FOV = 240$\times$240mm\textsuperscript{2}, matrix = 240$\times$240, slice thickness = 0.9mm, and voxel size = 0.9$\times$0.9$\times$0.9mm\textsuperscript{3}.
\item T2-Weighted Structural Imaging: Parameters were set at TR=5000ms, TE=105ms, flip angle = 150°, field of view (FOV) = 220$\times$200mm\textsuperscript{2}, matrix = 448$\times$358, slice thickness = 3.0mm, and voxel size = 0.5$\times$0.5$\times$3.0mm\textsuperscript{3}.
\item T2 FLAIR 3D Structural Imaging: Parameters were set at TR = 5000ms, TE = 387ms, FOV = 230$\times$230mm\textsuperscript{2}, matrix = 256$\times$256, slice thickness = 0.9mm, and voxel size = 0.9$\times$0.9$\times$0.9mm\textsuperscript{3}.
\end{itemize}

\begin{figure}
    \centering
    \includegraphics[width=1\linewidth]{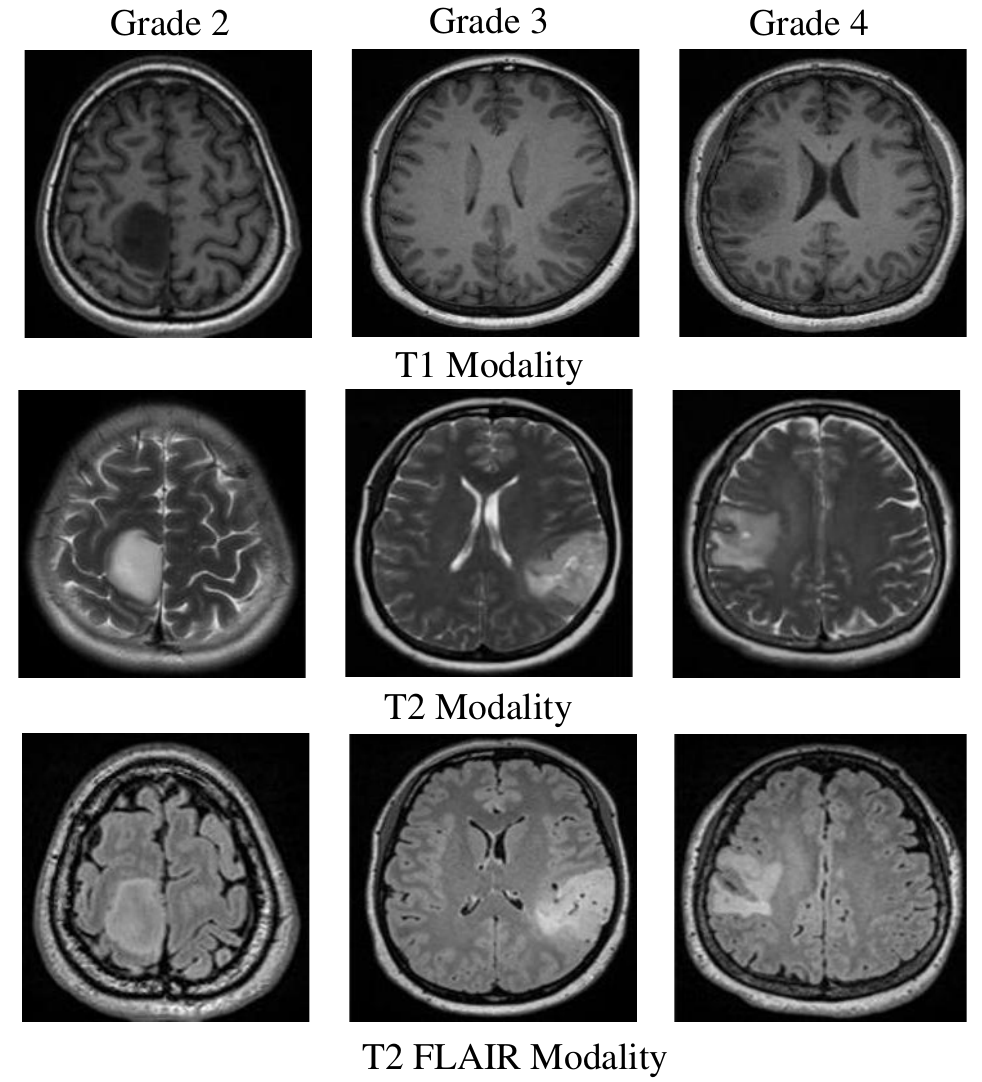}
    \caption{The figure presents MRI scans showcasing different modalities (T1, T2, and T2 FLAIR) across various glioma grades.}
    \label{fig:private}
\end{figure}

We gathered three distinct datasets from our patient cohort, each representing a different imaging modality. These include 100 MRIs in the T1 modality, 101 MRIs in T2, and 36 MRIs in T2 FLAIR. While we could not obtain all three modalities for each patient, the data we collected was meticulously organized by case number, and a radiologist with 8 years of experience (L.M) provided voxel-level tumor annotations for each image. To address potential discrepancies in image labels and voxel sizes across the datasets, we resampled both data and labels collectively to ensure uniformity and alignment for all patients. Our study focused on 2D axial slices extracted from the 3D data, specifically those where the tumor's Region of Interest (ROI), indicated by a binary mask, exceeded 30 pixels in size. These slices formed the basis of our three datasets, which were suitable for the 2D input requirements of our models. We split the datasets into training and test sets on a patient level to prevent data leakage~\cite{thibeau2022clinicadl}, maintaining a 7:3 ratio. Within the training dataset, we divided it into a training and validation set using an 8:2 ratio.
The example images in Figure~\ref{fig:private} illustrate the various modalities and tumor grades in our clinical experimental dataset.

\subsection{Public datasets}
\begin{figure*}
    \centering
    \includegraphics[width=1\linewidth]{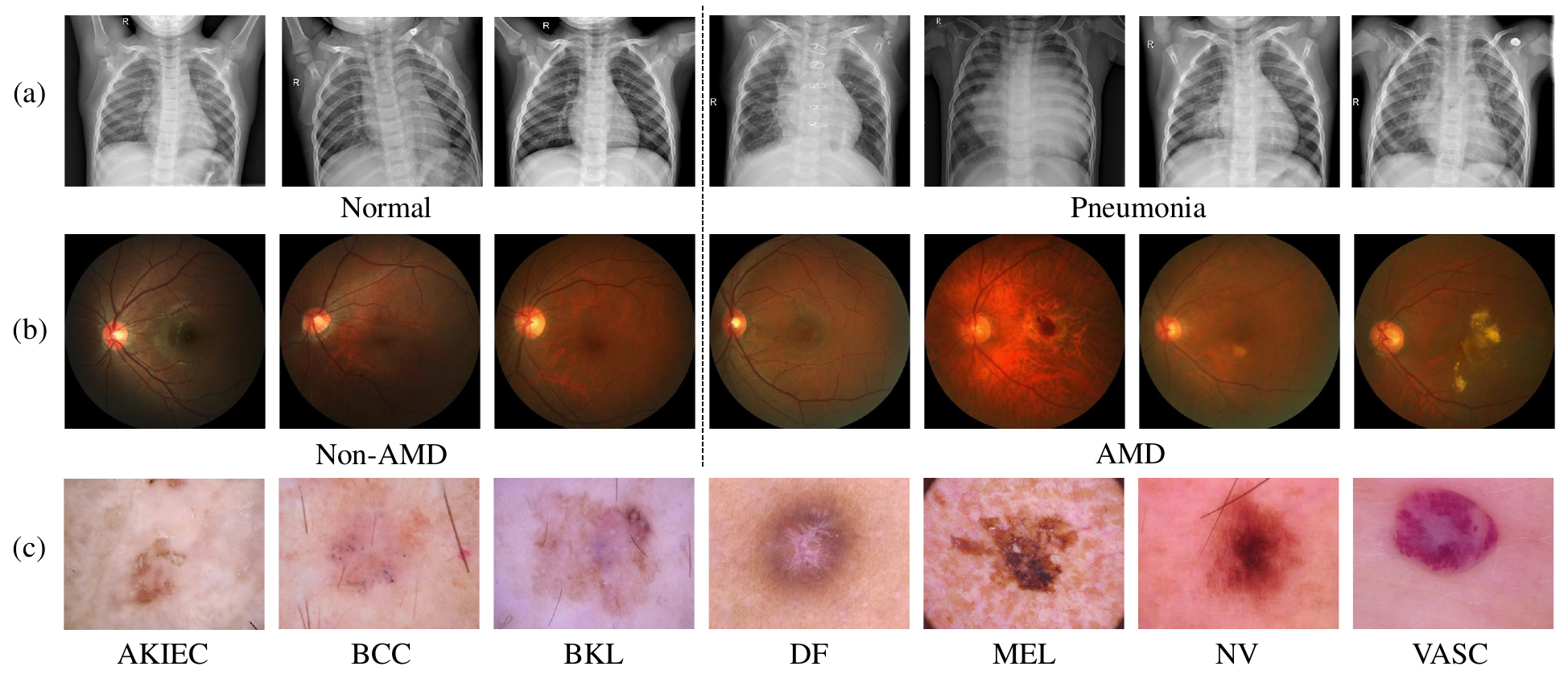}
    \caption{The examples from three public datasets include: (a) a Chest X-Ray dataset used for binary classification of pneumonia and normal cases, (b) the iChallenge-AMD dataset aimed at detecting age-related macular degeneration in eye fundus images, and (c) the HAM10000 dataset for classifying skin cancer in dermoscopy images.}
    \label{fig:public}
\end{figure*}

To support our research, we also used three public datasets, including the classification of chest X-rays, eye funds, and skin dermoscopic images. The example images for these datasets can be seen in Figure~\ref{fig:public}
Detailed data description is shown in Table~\ref{tab:data_split} and Table~\ref{tab:data_split_categories}.

\paragraph{Chest X-Ray dataset~\cite{chestxray_kermany2018identifying}} consists of patients aged 1 to 5 years at the Guangzhou Women and Children's Medical Center (GWCMC), and the data were evaluated and examined by three experts. The dataset is divided into 3 folders (training set: train, test set: test, validation set: val) and contains sub-folders for each image category (pneumonia/normal). The data set has 5840 X-ray images (4265 pneumonia images and 1575 normal images). We used the same train/test division, which contains 5216 images for training and 624 images for testing.

\paragraph{iChallenge-AMD dataset~\cite{eye_orlando2020refuge}} it consisted of retinal fundus images of age-related macular degeneration (AMD) from Chinese patients, with 77\% of non-AMD subjects (89 AMD images) and 23\% of AMD subjects (311 non-AMD images). Each image contains glaucoma/non-glaucoma labels assigned based on comprehensive clinical evaluation results. The images were collected by a Zeiss Visucam 500 funding camera (2124 $\times$ 2056 pixels) and a Canon CR-2 unit (1634 $\times$ 1634 pixels). This dataset label is assigned based on comprehensive clinical assessment. In this study, the training set has 245 images, and the test set has 155 images.

\paragraph{HAM10000 dataset~\cite{skin_tschandl2018ham10000}} it contains 11,513 dermoscopic images from the Department of Dermatology, Medical University of Vienna, Austria, and the Skin Cancer Clinic, Queensland, Australia, collected over 20 years in seven categories: 370 images of actinic keratoses and intraepithelial carcinomas/ Bowen's disease (AKIEC), 607 images of basal cell carcinomas (BCC), 1316 images of benign keratosis pilaris-like lesions (solar nevus/seborrheic keratosis and lichen planus-like keratosis, BKL), 159 images of dermatofibroma (DF), 1,286 images of melanoma (MEL), 6703 images of melanocytic nevus (NV), and 177 images of vascular lesions (hemangiomas, angiokeratomas, pyogenic granulomas, and hemorrhages, VASC). The images were collected using a film camera, MoleMax HD digital dermoscopy system (1872 x 1053 pixels), DermLite Fluid, and DermLite DL3 device. We keep the train/test division, consisting of 10015 images for training and 1512 for testing. 

\begin{table}
\centering
\caption{The data distributions of private and public datasets. Note that the private datasets are in 3D; we split the dataset at the patient level. \#C means the number of categories.}
\label{tab:data_split}
\begin{tabular}{|c|c|c|c|c|c|c|} 
\hline
\multirow{2}{*}{Dataset} & \multirow{2}{*}{Source}  & \multicolumn{2}{c|}{Patients}       & \multicolumn{2}{c|}{Slices} & \multirow{2}{*}{\#C}  \\ 
\cline{3-6}
                         &                          & Train            & Test             & Train & Test                &                       \\ 
\hline
T1 MRI                   & \multirow{3}{*}{Private} & 70               & 30               & 3822  & 1613                & 3                     \\ 
\cline{1-1}\cline{3-7}
T2 MRI                   &                          & 70               & 31               & 1061  & 457                 & 3                     \\ 
\cline{1-1}\cline{3-7}
T2 FLAIR MRI             &                          & 25               & 11               & 1501  & 576                 & 3                     \\ 
\hline
Chest X-Ray              & \multirow{3}{*}{Public}  & - & - & 5216  & 624                 & 2                     \\ 
\cline{1-1}\cline{3-7}
iChallenge-AMD           &                          & - & - & 245   & 155                 & 2                     \\ 
\cline{1-1}\cline{3-7}
HAM10000                 &                          & - & - & 10015 & 1512                & 7                     \\
\hline
\end{tabular}
\end{table}

\begin{table}[h!]
\caption{Details the quantity of data within each specific category of both the private and public datasets for train and test sets.}\label{tab:data_split_categories}
\centering
\begin{tabular}{|c|c|c|c|}
\hline
\multirow{2}{*}{Dataset} & \multirow{2}{*}{Categories} & \multicolumn{2}{c|}{Slices} \\ \cline{3-4}
                         &                           & Train & Test  \\ \hline
\multirow{3}{*}{T1 MRI}  & Grade2                    & 1392  & 510  \\ \cline{2-4} 
                         & Grade3                    & 1198  & 457  \\ \cline{2-4} 
                         & Grade4                    & 1232  & 646  \\ \hline
\multirow{3}{*}{T2 MRI}  & Grade2                    & 404   & 179  \\ \cline{2-4} 
                         & Grade3                    & 330   & 139  \\ \cline{2-4} 
                         & Grade4                    & 327   & 139  \\ \hline
\multirow{3}{*}{T2 FLAIR MRI} & Grade2               & 532   & 196   \\ \cline{2-4} 
                         & Grade3                    & 491   & 179   \\ \cline{2-4} 
                         & Grade4                    & 478   & 201  \\ \hline
\multirow{2}{*}{Chest X-Ray} & Normal                & 1341  & 234   \\ \cline{2-4} 
                         & Pneumonia                 & 3875  & 390   \\ \hline
\multirow{2}{*}{iChallenge-AMD}  & AMD               & 62    & 27      \\ \cline{2-4} 
                         & Non-AMD                   & 183   & 128    \\ \hline
\multirow{7}{*}{HAM10000} & AKIEC                    & 327   & 43    \\ \cline{2-4} 
                         & BCC                       & 514   & 93     \\ \cline{2-4} 
                         & BKL                       & 1099  & 217     \\ \cline{2-4} 
                         & DF                        & 115   & 44    \\ \cline{2-4} 
                         & MEL                       & 1113  & 171      \\ \cline{2-4} 
                         & NV                        & 6705  & 909     \\ \cline{2-4} 
                         & VASC                      & 142   & 35     \\ \hline

\end{tabular}
\end{table}

\section{Experiment details}
The above public and clinical datasets have undergone the same data preprocessing: 1) Resize the input image to 256$\times$256 pixels. 2) Perform a center crop operation on the image to 224$\times$224 pixels. 3) Normalize images by mean and standard deviation. 
Our models were trained for 100 epochs with a batch size of 32, using the Adam optimizer~\cite{kingma2014adam} with a learning rate of \(1 \times 10^{-4}\). We employed the cross-entropy loss function throughout the training process.
The implementation used the PyTorch framework~\cite{paszke2019pytorch}. We used pre-trained weights from ImageNet, available via TorchVision~\cite{marcel2010torchvision} for initializing the models.
For the experiments involving DINOv2, weights were downloaded from the official website\footnote{\url{https://github.com/facebookresearch/DINOv2}}.
Our experiments included different versions of the DINOv2 model, ranging from small (DINOv2-s) to large (DINOv2-l). However, due to GPU capacity limitations, we could not run the DINOv2-l model on our private dataset.
In our experiments, we kept the original architecture of the models, making adjustments only to the final classification layer. These modifications were implemented to address the varying number of classification categories pertinent to each specific task. Furthermore, the study systematically examines the effect of the freezing mechanism in the context of transfer learning.

As detailed before, all classifications and assessments were conducted at the patient level to prevent data leakage~\cite{thibeau2022clinicadl}. While the code is publicly accessible for reproducibility~\cite{colliot2023reproducibility}, the clinical data remains confidential to uphold patient privacy.
When choosing performance metrics for our analysis, we specifically employed accuracy, precision, recall, and F1-score, all of which are averaged using the weighted method. For each metric, we report its mean value over the independent test set as well as the corresponding 95\% confidence interval (CI) computed using bootstrap. 
The code for reproducing our experiments is available at the following link: \url{https://github.com/GuanghuiFU/medical_dino_eval}.

\section{Results and discussion}




Our experimental analysis focused on comparing the performance of DINOv2 and ImageNet pre-trained models (VGG, ResNet, DenseNet) across both private and public datasets, with a particular emphasis on the impact of the freezing mechanism. 
The comparative results of these experiments are presented in Table~\ref{tab:exp_private} for brain tumor grading tasks in the private dataset and in Table~\ref{tab:exp_public} for the analyses involving the public datasets. This section presents the key results and discusses the main insights about the observed models' performance. 

Regarding the freezing mechanism, our study revealed that the impact of freezing layers in ImageNet pre-trained models, such as VGG, ResNet, and DenseNet, exhibits significant variability in the model performance, which highlights the complexity of determining the optimal strategy for leveraging pre-trained models in new tasks. It also suggests that the decision to freeze layers within these architectures cannot be universally applied but instead necessitates a refined approach (i.e., hyperparameter refinement) tailored to the characteristics of the specific dataset and the requirements of the task at hand. 

In contrast, applying the freezing mechanism to DINOv2 models consistently improved performance across private and public datasets. This uniformity indicates that the DINOv2 architecture is particularly receptive to freezing layers, benefiting from its intrinsic stability and efficiently leveraging pre-existing learned features. Such findings suggest that DINOv2 models possess characteristics that make them more adaptable to the preservation of learned representations, enhancing their ability to generalize across different tasks and datasets without extensive retraining.

When considering the performance in the publicly selected datasets, DINOv2 models outperformed the traditional ImageNet pre-trained models. These results suggest that the DINOv2 models' architecture is robust enough to handle different types of public medical data independently of the freezing mechanism. In addition, an interesting finding was the high performance of DINOv2 models (max of 86.73\% F1 score) in the eye funds dataset (i.e., iChallenge-AMD), which had a relatively lower amount of images than the other publicly available datasets. This increased performance occurs when applying the freezing mechanism, demonstrating DINOv2's capability to transfer the learned features to limited data effectively, an essential quality for tasks with constrained data availability.

On the other hand, in our private datasets, DINOv2 models exhibited slightly weaker and sometimes comparable performance than the ImageNet pre-trained models. All models, however, achieved quantitatively low results, with the max F1 score of 49.74\% in the T2 dataset using the ResNet50. These results suggest that while DINOv2 models are robust in more generalized settings, they may require further tuning or adaptation to excel in specialized, private datasets. Especially when the classes in the dataset do not present apparent visual differences, as is the case between Grades 3 and 4 for T1, T2, and T2FLAIR modalities (as observed in Figure~\ref{fig:private}). Then, these observations highlight the necessity of considering the dataset-specific features (like diversity, volume, and label distribution)
in the model selection process.

Interestingly, we also found that larger versions of DINOv2 did not necessarily outperform their smaller counterparts. This observation suggests that increased model complexity in DINOv2 does not always correlate with enhanced performance, emphasizing the need for balance between model size and task complexity in addition to the data distribution and characteristics considerations.

Our study, centered on pre-trained models, was designed to facilitate an equitable comparison across different architectures. However, this choice inherently limited our exploration of each model's intrinsic capabilities. Future research should consider a more holistic approach, potentially incorporating models trained from scratch, to uncover alternative performance characteristics and applicability across different tasks and image modalities.

The observed performance between different versions of DINOv2 prompts an essential question regarding the optimal balance between model complexity and computational efficiency. Future studies could also focus on identifying the most effective model size for specific tasks, potentially leading to more computationally economical yet practical solutions.

\begin{table*}[ht]
\centering
\caption{The classification task performance across three modalities of private datasets is presented as the mean with a 95\% bootstrap confidence interval. Metrics such as precision, recall, and F1-score are reported as weighted averages.}
\label{tab:exp_private}
\begin{tabular}{|c|c|c|c|c|c|c|} 
\hline
Dataset                       & Model                              & Freeze & Accuracy             & Precision            & Recall               & F1 Score              \\ 
\hline
\multirow{10}{*}{T1}       & \multirow{2}{*}{VGG16}       & N     & 41.29 [41.25, 41.32] & 38.90 [38.86, 38.93] & 41.29 [41.25, 41.32] & 39.45 [39.41, 39.48]  \\ 
\cline{3-7}
                           &                                    & Y    & 46.94 [46.91, 46.97] & 45.26 [45.23, 45.30] & 46.94 [46.91, 46.97] & 45.57 [45.53, 45.60]  \\ 
\cline{2-7}
                           & \multirow{2}{*}{ResNet50}     & N     & 38.89 [38.86, 38.92] & 38.24 [38.20, 38.27] & 38.89 [38.86, 38.92] & 38.48 [38.44, 38.51]  \\ 
\cline{3-7}
                           &                                    & Y    & 38.82 [38.78, 38.85] & 38.95 [38.92, 38.99] & 38.82 [38.78, 38.85] & 38.86 [38.83, 38.90]  \\ 
\cline{2-7}
                           & \multirow{2}{*}{DenseNet121} & N     & 46.39 [46.36, 46.43] & 46.45 [46.42, 46.49] & 46.39 [46.36, 46.43] & 46.39 [46.35, 46.42]  \\ 
\cline{3-7}
                           &                                    & Y    & 32.00 [31.97, 32.03] & 32.20 [32.17, 32.23] & 32.00 [31.97, 32.03] & 32.00 [31.96, 32.03]  \\

\hhline{|~|======|} 
                           & \multirow{2}{*}{DINOv2-s} & N     & 27.78 [27.75, 27.81] & 29.90 [29.87, 29.94] & 27.78 [27.75, 27.81] & 28.48 [28.45, 28.51]  \\ 
\cline{3-7}
                           &                                    & Y    & 38.48 [38.44, 38.51] & 38.38 [38.34, 38.41] & 38.48 [38.44, 38.51] & 38.08 [38.05, 38.12]  \\ 
\cline{2-7}
                           & \multirow{2}{*}{DINOv2-b} & N     & 26.76 [26.73, 26.79] & 28.50 [28.46, 28.53] & 26.76 [26.73, 26.79] & 27.44 [27.41, 27.47]  \\ 
\cline{3-7}
                           &                                    & Y    & 40.68 [40.65, 40.72] & 42.00 [41.97, 42.04] & 40.68 [40.65, 40.72] & 41.00 [40.97, 41.04]  \\ 
\hhline{=======}

\multirow{10}{*}{T2}       & \multirow{2}{*}{VGG16}       & N     & 44.66 [44.59, 44.72] & 47.96 [47.89, 48.03] & 44.66 [44.59, 44.72] & 44.76 [44.69, 44.82]  \\ 
\cline{3-7}
                           &                                    & Y    & 38.51 [38.45, 38.58] & 37.80 [37.73, 37.86] & 38.51 [38.45, 38.58] & 37.98 [37.91, 38.04]  \\ 
\cline{2-7}
                           & \multirow{2}{*}{ResNet50}     & N     & 50.11 [50.05, 50.18] & 51.20 [51.14, 51.28] & 50.11 [50.05, 50.18] & 49.74 [49.67, 49.81]  \\ 
\cline{3-7}
                           &                                    & Y    & 37.24 [37.18, 37.30] & 37.67 [37.60, 37.73] & 37.24 [37.18, 37.30] & 37.12 [37.05, 37.18]  \\ 
\cline{2-7}
                           & \multirow{2}{*}{DenseNet121} & N     & 40.44 [40.38, 40.50] & 40.63 [40.57, 40.70] & 40.44 [40.38, 40.50] & 40.44 [40.37, 40.50]  \\ 
\cline{3-7}
                           &                                    & Y    & 37.44 [37.38, 37.50] & 37.35 [37.29, 37.42] & 37.44 [37.38, 37.50] & 37.02 [36.96, 37.09]  \\ 
\hhline{|~|======|} 
                           & \multirow{2}{*}{DINOv2-s} & N     & 32.39 [32.33, 32.45] & 32.13 [32.07, 32.19] & 32.39 [32.33, 32.45] & 31.81 [31.75, 31.87]  \\ 
\cline{3-7}
                           &                                    & Y    & 43.80 [43.74, 43.87] & 45.44 [45.37, 45.51] & 43.80 [43.74, 43.87] & 44.08 [44.02, 44.15]  \\ 
\cline{2-7}
                           & \multirow{2}{*}{DINOv2-b} & N     & 30.85 [30.79, 30.91] & 30.78 [30.72, 30.85] & 30.85 [30.79, 30.91] & 30.62 [30.56, 30.69]  \\ 
\cline{3-7}
                           &                                    & Y    & 42.70 [42.63, 42.76] & 46.74 [46.66, 46.81] & 42.70 [42.63, 42.76] & 42.26 [42.19, 42.32]  \\ 
\hhline{=======}
\multirow{10}{*}{T2 FLAIR} & \multirow{2}{*}{VGG16}       & N     & 27.63 [27.58, 27.68] & 22.67 [22.60, 22.74] & 27.63 [27.58, 27.68] & 20.74 [20.69, 20.79]  \\ 
\cline{3-7}
                           &                                    & Y    & 31.24 [31.19, 31.30] & 27.63 [27.57, 27.69] & 31.24 [31.19, 31.30] & 28.21 [28.15, 28.26]  \\ 
\cline{2-7}
                           & \multirow{2}{*}{ResNet50}     & N     & 40.29 [40.23, 40.35] & 37.01 [36.95, 37.08] & 40.29 [40.23, 40.35] & 37.06 [37.01, 37.12]  \\ 
\cline{3-7}
                           &                                    & Y    & 28.98 [28.93, 29.04] & 27.02 [26.96, 27.08] & 28.98 [28.93, 29.04] & 25.47 [25.42, 25.53]  \\ 
\cline{2-7}
                           & \multirow{2}{*}{DenseNet121} & N     & 36.96 [36.90, 37.02] & 37.58 [37.51, 37.66] & 36.96 [36.90, 37.02] & 32.65 [32.59, 32.71]  \\ 
\cline{3-7}
                           &                                    & Y    & 36.62 [36.57, 36.68] & 36.95 [36.89, 37.02] & 36.62 [36.57, 36.68] & 34.23 [34.17, 34.28]  \\ 
\hhline{|~|======|} 
                           & \multirow{2}{*}{DINOv2-s} & N     & 37.63 [37.57, 37.68] & 37.10 [37.04, 37.16] & 37.63 [37.57, 37.68] & 34.86 [34.80, 34.92]  \\ 
\cline{3-7}
                           &                                    & Y    & 39.80 [39.74, 39.86] & 42.73 [42.66, 42.80] & 39.80 [39.74, 39.86] & 37.02 [36.95, 37.08]  \\ 
\cline{2-7}
                           & \multirow{2}{*}{DINOv2-b} & N     & 35.23 [35.17, 35.28] & 33.67 [33.61, 33.73] & 35.23 [35.17, 35.28] & 33.40 [33.35, 33.46]  \\ 
\cline{3-7}
                           &                                    & Y    & 39.10 [39.04, 39.15] & 47.04 [46.96, 47.12] & 39.10 [39.04, 39.15] & 35.37 [35.31, 35.42]  \\
\hline
\end{tabular}
\end{table*}

\begin{table*}
\centering
\caption{The classification task performance across three public datasets are presented as the mean with a 95\% bootstrap confidence interval. Metrics such as precision, recall, and F1-score are reported as weighted averages.}
\label{tab:exp_public}
\begin{tabular}{|c|c|c|c|c|c|c|} 
\hline
Dataset                             & Model                              & Freeze & Accuracy             & Precision            & Recall               & F1 Score              \\ 
\hline
\multirow{12}{*}{Chest X-ray}    & \multirow{2}{*}{VGG16}       & N     & 81.40 [81.35, 81.44] & 84.37 [84.34, 84.40] & 81.40 [81.35, 81.44] & 79.82 [79.77, 79.87]  \\ 
\cline{3-7}
                                 &                                    & Y    & 78.87 [78.82, 78.91] & 83.11 [83.08, 83.15] & 78.87 [78.82, 78.91] & 76.49 [76.44, 76.55]  \\ 
\cline{2-7}
                                 & \multirow{2}{*}{ResNet50}     & N     & 75.29 [75.24, 75.34] & 81.16 [81.13, 81.19] & 75.29 [75.24, 75.34] & 71.53 [71.47, 71.59]  \\ 
\cline{3-7}
                                 &                                    & Y    & 76.63 [76.59, 76.68] & 79.30 [79.26, 79.34] & 76.63 [76.59, 76.68] & 74.22 [74.16, 74.28]  \\ 
\cline{2-7}
                                 & \multirow{2}{*}{DenseNet121} & N     & 73.24 [73.20, 73.29] & 80.34 [80.31, 80.37] & 73.24 [73.20, 73.29] & 68.40 [68.34, 68.46]  \\ 
\cline{3-7}
                                 &                                    & Y    & 79.98 [79.93, 80.02] & 83.40 [83.37, 83.44] & 79.98 [79.93, 80.02] & 78.04 [77.99, 78.10]  \\ 
\hhline{|~|======|} 
                                 & \multirow{2}{*}{DINOv2-s} & N     & 80.90 [80.86, 80.95] & 82.85 [82.81, 82.89] & 80.90 [80.86, 80.95] & 79.54 [79.49, 79.59]  \\ 
\cline{3-7}
                                 &                                    & Y    & 83.82 [83.77, 83.86] & 86.97 [86.94, 86.99] & 83.82 [83.77, 83.86] & 82.54 [82.50, 82.59]  \\ 
\cline{2-7}
                                 & \multirow{2}{*}{DINOv2-b} & N     & 77.38 [77.33, 77.42] & 80.29 [80.25, 80.33] & 77.38 [77.33, 77.42] & 75.07 [75.01, 75.12]  \\ 
\cline{3-7}
                                 &                                    & Y    & 82.84 [82.80, 82.88] & 86.13 [86.11, 86.16] & 82.84 [82.80, 82.88] & 81.41 [81.36, 81.46]  \\ 
\cline{2-7}
                                 & \multirow{2}{*}{DINOv2-l} & N     & 78.49 [78.44, 78.54] & 80.95 [80.91, 80.99] & 78.49 [78.44, 78.54] & 76.55 [76.50, 76.61]  \\ 
\cline{3-7}
                                 &                                    & Y    & 83.00 [82.96, 83.04] & 86.45 [86.43, 86.48] & 83.00 [82.96, 83.04] & 81.56 [81.51, 81.61]  \\ 
\hhline{=======}
\multirow{12}{*}{iChallenge-AMD} & \multirow{2}{*}{VGG16}       & N     & 80.67 [80.59, 80.76] & 73.86 [73.72, 74.01] & 80.67 [80.59, 80.76] & 75.67 [75.56, 75.79]  \\ 
\cline{3-7}
                                 &                                    & Y    & 83.21 [83.12, 83.29] & 79.91 [79.66, 80.15] & 83.21 [83.12, 83.29] & 76.18 [76.06, 76.30]  \\ 
\cline{2-7}
                                 & \multirow{2}{*}{ResNet50}     & N     & 80.66 [80.57, 80.74] & 71.95 [71.79, 72.10] & 80.66 [80.57, 80.74] & 74.76 [74.64, 74.87]  \\ 
\cline{3-7}
                                 &                                    & Y    & 82.53 [82.45, 82.62] & 68.21 [68.07, 68.35] & 82.53 [82.45, 82.62] & 74.67 [74.55, 74.78]  \\ 
\cline{2-7}
                                 & \multirow{2}{*}{DenseNet121} & N     & 81.19 [81.10, 81.28] & 79.00 [78.88, 79.10] & 81.19 [81.10, 81.28] & 79.60 [79.49, 79.69]  \\ 
\cline{3-7}
                                 &                                    & Y    & 83.24 [83.16, 83.32] & 80.29 [80.11, 80.47] & 83.24 [83.16, 83.32] & 77.23 [77.12, 77.34]  \\ 
\hhline{|~|======|} 
                                 & \multirow{2}{*}{DINOv2-s} & N     & 76.76 [76.66, 76.85] & 69.30 [69.16, 69.43] & 76.76 [76.66, 76.85] & 72.50 [72.38, 72.61]  \\ 
\cline{3-7}
                                 &                                    & Y    & 87.12 [87.05, 87.20] & 86.12 [86.03, 86.21] & 87.12 [87.05, 87.20] & 85.51 [85.42, 85.60]  \\ 
\cline{2-7}
                                 & \multirow{2}{*}{DINOv2-b} & N     & 79.29 [79.19, 79.38] & 76.31 [76.19, 76.42] & 79.29 [79.19, 79.38] & 77.30 [77.19, 77.40]  \\ 
\cline{3-7}
                                 &                                    & Y    & 87.16 [87.09, 87.24] & 86.71 [86.62, 86.79] & 87.16 [87.09, 87.24] & 86.73 [86.65, 86.81]  \\ 
\cline{2-7}
                                 & \multirow{2}{*}{DINOv2-l} & N     & 81.88 [81.80, 81.97] & 81.58 [81.48, 81.67] & 81.88 [81.80, 81.97] & 81.60 [81.51, 81.68]  \\ 
\cline{3-7}
                                 &                                    & Y    & 87.73 [87.65, 87.80] & 88.12 [88.04, 88.19] & 87.73 [87.65, 87.80] & 85.19 [85.09, 85.28]  \\ 
\hhline{=======}
\multirow{12}{*}{HAM10000}       & \multirow{2}{*}{VGG16}       & N     & 64.83 [64.80, 64.87] & 61.07 [61.03, 61.12] & 64.83 [64.80, 64.87] & 61.97 [61.93, 62.01]  \\ 
\cline{3-7}
                                 &                                    & Y    & 67.35 [67.32, 67.38] & 64.22 [64.18, 64.26] & 67.35 [67.32, 67.38] & 64.58 [64.54, 64.61]  \\ 
\cline{2-7}
                                 & \multirow{2}{*}{ResNet50}     & N     & 65.97 [65.94, 66.01] & 62.05 [62.01, 62.09] & 65.97 [65.94, 66.01] & 63.48 [63.44, 63.52]  \\ 
\cline{3-7}
                                 &                                    & Y    & 59.92 [59.88, 59.96] & 50.07 [50.01, 50.12] & 59.92 [59.88, 59.96] & 51.51 [51.46, 51.55]  \\ 
\cline{2-7}
                                 & \multirow{2}{*}{DenseNet121} & N     & 68.88 [68.84, 68.91] & 66.55 [66.51, 66.59] & 68.88 [68.84, 68.91] & 67.17 [67.13, 67.20]  \\ 
\cline{3-7}
                                 &                                    & Y    & 66.12 [66.08, 66.15] & 61.26 [61.21, 61.30] & 66.12 [66.08, 66.15] & 62.02 [61.98, 62.06]  \\ 
\hhline{|~|======|} 
                                 & \multirow{2}{*}{DINOv2-s} & N     & 69.54 [69.51, 69.57] & 69.43 [69.39, 69.47] & 69.54 [69.51, 69.57] & 68.92 [68.89, 68.96]  \\ 
\cline{3-7}
                                 &                                    & Y    & 77.33 [77.30, 77.36] & 77.13 [77.10, 77.16] & 77.33 [77.30, 77.36] & 76.70 [76.67, 76.73]  \\ 
\cline{2-7}
                                 & \multirow{2}{*}{DINOv2-b} & N     & 69.70 [69.67, 69.73] & 69.10 [69.06, 69.13] & 69.70 [69.67, 69.73] & 69.02 [68.99, 69.06]  \\ 
\cline{3-7}
                                 &                                    & Y    & 78.78 [78.75, 78.81] & 78.17 [78.14, 78.20] & 78.78 [78.75, 78.81] & 78.02 [77.99, 78.06]  \\ 
\cline{2-7}
                                 & \multirow{2}{*}{DINOv2-l} & N     & 69.77 [69.74, 69.81] & 68.27 [68.23, 68.31] & 69.77 [69.74, 69.81] & 68.60 [68.56, 68.63]  \\ 
\cline{3-7}
                                 &                                    & Y    & 78.71 [78.68, 78.74] & 79.13 [79.10, 79.16] & 78.71 [78.68, 78.74] & 78.54 [78.51, 78.57]  \\
\hline
\end{tabular}
\end{table*}

\section{Conclusion}
This study conducted a comprehensive analysis of DINOv2 and ImageNet pre-trained models across various datasets, including public and private collections, focusing specifically on brain glioma grading tasks. This investigation yielded several key insights that contribute to the existing knowledge in machine learning and set a direction for future research.

Firstly, the study highlighted the significant influence of dataset characteristics on model performance. DINOv2 models excelled in public datasets, demonstrating their robustness and adaptability. However, in the more specialized context of private brain glioma datasets, they were outperformed by ImageNet pre-trained models. This finding emphasizes the need to consider dataset specificity when choosing models for particular applications.
The research also examined the impact of the freezing mechanism on model performance. DINOv2 models consistently showed improvement with freezing, while the response of ImageNet pre-trained models varied. This observation suggests that tailored training strategies could improve model performance, especially in transfer learning contexts.
Nevertheless, the study's focus on pre-trained models highlights a limitation: the lack of an examination study of models trained from scratch. Investigating such models could provide additional insights into their capabilities and adaptability.
This study set the basis for comparing classic deep learning approaches and foundation models and opens several paths for future research. There is a compelling need to test a broader range of tasks and datasets to corroborate and expand upon these findings. Further exploration into the optimal balance between model size and complexity, the effects of various training strategies, and the investigation of transfer learning and fine-tuning methods are vital for the progression of the field.

In summary, this study offers valuable perspectives on the performance of advanced deep learning models across diverse datasets. It highlights the crucial role of dataset characteristics and training strategies in achieving optimal model performance. These insights are pivotal for continuously evolving and refining machine learning models, particularly in their application to specialized and varied datasets.


\bibliographystyle{IEEEtran}
\bibliography{ref}
\end{document}